# Non-Hermitian glide-time symmetry

Li-Wei Wang[1, 2] and Jian-Hua Jiang[1, 2, 3, †]

[1]*School of Physical Sciences, University of Science and Technology of China, Hefei, 230026, China*

[2]*Suzhou Institute for Advanced Research, University of Science and Technology of China, Suzhou 215123, China*

[3]*School of Biomedical Engineering, Division of Life Sciences and Medicine, University of Science and Technology of China, Hefei 230026, China*

[†]Correspondence should be addressed to: jhjiang3@ustc.edu.cn

**Abstract**

Non-Hermitian systems, going beyond conventional Hermitian systems, have brought in intriguing concepts such as exceptional points and complex spectral topology as well as exotic phenomena such as non-Hermitian skin effects (NHSEs). However, previous studies on non-Hermitian systems predominantly focus on the properties of eigenstates, with rather limited discussions on non-Hermitian dynamic phenomena. Here, inspired by the celebrated success of the parity-time symmetry in non-Hermitian physics, we theoretically study a one-dimensional non-Hermitian system with glide-time reversal (*GT*) symmetry. We discover that the *GT* symmetry leads to unique physical properties and enables rich dynamic phenomena in non-Hermitian systems. Remarkably, we reveal the dynamic NHSEs that exhibit diverse behaviors across distinct dynamic phases, elucidating the richness of non-Hermitian dynamics. We establish the theoretical frameworks for understanding the rich non-Hermitian dynamic phenomena. We further show that the rich dynamic phases in the *GT*-symmetric systems enable the remarkable tuning of the dynamics in the bulk as well as at the edge boundaries. These include the directional wave propagation and amplification in the bulk, as well as the wave trapping and the dynamic patterns at the edge boundaries. With both the development in the theoretical framework and the study of the rich non-Hermitian dynamic phases, this work serves as a stepstone for future studies on non-Hermitian dynamics with a special emphasize on the pivotal role of the lattice symmetry.

# I . INTRODUCTION

Non-Hermitian systems [1,2] and their related effects have attracted widespread research interest, leading to numerous discoveries in electron systems [3,4], classical systems [5–7] and open quantum systems [8–11] that have significantly broadened the understanding on non-equilibrium properties of matter. The energy spectra of non-Hermitian systems can be complex, giving rise to non-Hermitian knot topology [12,13] and spectral topology [14–19], which are distinct from their Hermitian counterparts [20]. Additionally, in non-Hermitian systems, eigenstates exhibit mutual intertwining instead of orthogonality. These characteristics greatly advance our understanding of dynamics beyond the realm of energy conservation and lead to unprecedented phenomena such as the unidirectional invisibility [21] and the non-Hermitian state permutations [22]. Recently, it has been discovered that non-Hermitian lattice systems exhibit the presence of NHSEs, indicating the localization of all eigenstates at the boundaries [23–31]. Dealing with such systems necessitates a reexamination of the concept of Brillouin zone, which is a fundamental concept in condensed matter physics. Consequently, this leads to the introduction of generalized Brillouin zones (GBZs) [32].

To date, the majority of research on non-Hermitian systems has primarily focused on static eigenstate properties. The dynamic properties are essential for understanding phenomena in experiments. However, due to the existence of the exceptional points and other singularities, the eigenstates alone may be insufficient for studying the dynamics of non-Hermitian systems. In general, the understanding of non-Hermitian dynamics becomes challenging in the presence of complex energy spectra, resulting in intricate temporal dependence in the amplitudes of the eigenmodes. Moreover, non-Hermitian systems can exhibit boundary and defect scatterings that give rise to extraordinary phenomena beyond conventional dynamic understanding, such as wave self-healing [33] and inelastic boundary scatterings [34]. Recently, there has been a surge of study on non-Hermitian dynamics with remarkable findings such as chiral Zener tunneling [35], anharmonic Rabi oscillations [36], non-Bloch dynamics [37–39], dynamic NHSEs [34,40,41], non-Hermitian edge burst [42,43], and non-Hermitian wave self-acceleration [44], demonstrate that non-Hermitian dynamics can lead to phenomena extending beyond the static properties of eigenstates. These findings indicate that there are still much to explore in non-Hermitian dynamics where interesting phenomena are to be uncovered and fundamental elements are to be established. For instance, the important role of symmetry on non-Hermitian dynamics has not yet been clearly revealed. Another principal challenge in this emerging field lies in characterizing and classifying non-Hermitian dynamics by uncovering various dynamic phases and their transitions.

In this study, we illustrate the diverse non-Hermitian skin dynamics and emerging non-Hermitian dynamic phases within one-dimensional (1D) nonreciprocal double-chain configurations possessing *GT* symmetry. In our configurations, we connect two Su-Schrieffer-Heeger (SSH) chains with only the inter-chain couplings exhibiting nonreciprocity. By carefully adjusting the strength of these non-reciprocal couplings, we observe a rich variety of intriguing non-Hermitian dynamic phenomena. Additionally, we introduce novel techniques based on time evolution within the GBZs and within the eigenmodes under open boundary conditions (OBC) to comprehensively analyze and understand the versatile nature of these non-Hermitian dynamics.

The paper is organized as follows. In Sec. II, we introduce the non-Hermitian *GT* mode and conduct an analysis of its band topology and non-Hermitian topology under PBC. In Sec. III, We conducted OBC analysis for each phase, laying the foundation for subsequent dynamic analysis. In Sec. IV, we have presented the dynamic phase diagram of a non-Hermitian *GT* model and proposed a methodology for categorizing the distinct dynamics. Then, The origin of the amplification of non-Hermitian dynamics was proposed. In Sec. V. A, we employ Z transformation to analyse and comprehend the direction of dynamic NHSE, and provide illustrative analyses of three dynamic phases and the Hermitian phase in Secs. V. B, respectively. Finally, our conclusions are presented in Sec. V I.

## II. PHASE DIAGRAM ANALYSIS UNDER PBC

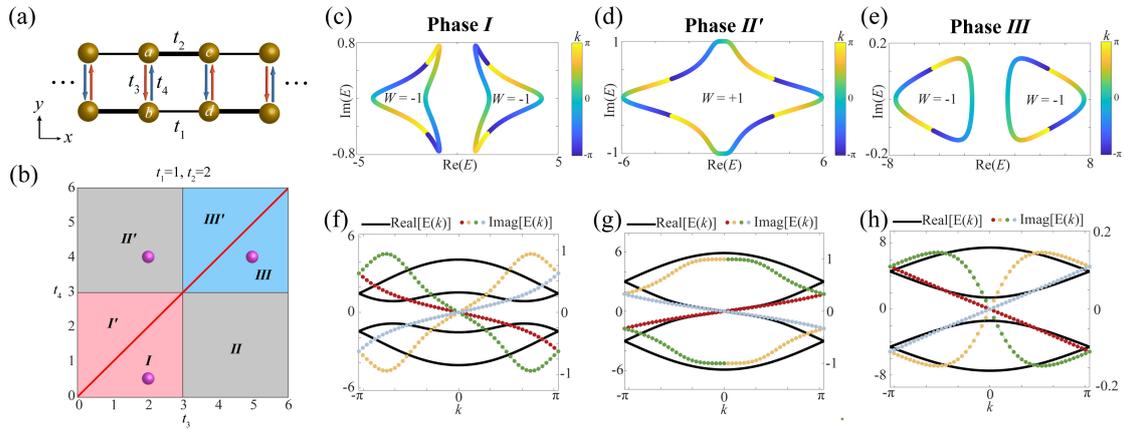

FIG. 1. Non-Hermitian *GT* model and eigenmodes phase diagram. (a) The structure of the non-Hermitian *GT* model. Each unit cell is composed of four sites $a,b,c,d$. $t_1$ and $t_2$

represent reciprocal couplings, while $t_3$ and $t_4$ denote non-reciprocal couplings. (b) Eigenmodes phase diagram. The red line represents the Hermitian condition, indicating $t_3 = t_4$. The regions in pink, blue, and gray, which are determined by the calculated Zak phase, correspond to the non-trivial, trivial, and gapless phases respectively. The phases $O$ ($O=\{$ I, II, III$\}$) exhibit NHSEs towards the left, whereas phase $O'$ exhibit NHSEs in the rightward direction. The purple spheres represent three selected parameter sets $(t_3, t_4)$: phase I (2,0.5), phase II′(2,4), phase III(5,4). (c)-(e) The complex energy spectra under periodic boundary conditions(PBC) for three cases in (b). The energy calculated for different wave vectors $k$ is represented in different colors. The winding number demonstrates within each energy loop. (f)-(h) Energy bands $E(k)$ for three cases in (b). The real parts of $E(k)$ are denoted by black curves and associated with the left scale axis. The imaginary parts of four energy bands are marked in red, orange, green and dusty blue, respectively, and associated with the right scale axis.

The non-Hermitian $GT$ model is pictorially shown in Fig.1(a), comprising of two SSH chains connected by non-reciprocal couplings. These two SSH chains are obtained by applying a mirror-symmetry operation to a designated SSH chain, followed by a translation of half the lattice constant along the $x$-direction. The Hamiltonian is given by

$$H = \sum_n \left( t_2 \hat{a}_n^\dagger \hat{c}_n + t_1 \hat{a}_{n+1}^\dagger \hat{c}_n + t_1 \hat{b}_n^\dagger \hat{d}_n + t_2 \hat{b}_{n+1}^\dagger \hat{d}_n + h.c. \right)$$

$$+ \sum_n (t_4 \hat{a}_n^\dagger \hat{b}_n + t_3 \hat{b}_n^\dagger \hat{a}_n + t_3 \hat{c}_n^\dagger \hat{d}_n + t_4 \hat{d}_n^\dagger \hat{c}_n) \qquad (1)$$

where $t_1$, $t_2$, $t_3$, and $t_4$ are real. The model possesses a glide symmetry $GH(k)G^{-1} = H(k)$, where the glide operation $G = e^{\frac{ik}{2}}(\cos\frac{k}{2}\sigma_1\tau_1 + \sin\frac{k}{2}\sigma_2\tau_1)$ corresponds to the transformation $(x,y) \rightarrow (x+\frac{1}{2}, -y)$, with $\sigma_i$ and $\tau_i$ representing Pauli matrixes. Due to the purely real nature of all couplings, the system inherently exhibits time-reversal symmetry $TH(k)T^{-1} = H(-k)$. Consequently, the model possesses both the glide symmetry $G$ and the time-reversal symmetry $T$.

In this study, the complex crystal symmetries (i.e., $GT$ symmetries) play a crucial role in phase enrichment, leading to a wider variety of topological and non-Hermitian phases. Here, we place emphasis on both the topological windings of eigenmodes in PBC and the complex energy spectra under PBC. The former serves as an indicator for the emergence of topological edge modes, while the latter provides insights into the presence or absence of NHSE. First, to characterize the band topology, we employ the complex Zak phase[45], expressing it as

$$\theta_Z = i \oint_{-\pi}^{\pi} dk \, \langle \psi_L(k) | \partial_k | \psi_R(k) \rangle, \qquad (2)$$

where $\psi_L(k)$ and $\psi_R(k)$ are the left and right eigenvectors of the Bloch Hamiltonian. The phase diagram is presented in Fig. 1(b) for $t_1 = 1$ and $t_2 = 2$. Notably, the phase boundaries are determined by $t_1 = 3$ or $t_2 = 3$. The pink (phase $I(I')$), blue (phase $II(II')$), and gray (phase $III(III')$) regions, which are determined by the calculated Zak phase, correspond to the non-trivial ($\theta_Z = \pi$), trivial ($\theta_Z = 0$), and gapless phases respectively. It is worth mentioning that compared to the non-Hermitian SSH model[32], the non-Hermitian *GT* model offers a richer variety of phases, including a gapless region spanning multiple parameter values rather than being limited to a single point.

Subsequently, the occurrence of NHSE can be determined either by employing the GBZ or calculating the winding number from the PBC complex energy spectra. The winding number for the Bloch Hamiltonian is defined as[24]:

$$W_E = \int_0^{2\pi} \frac{dk}{2\pi i} \frac{d}{dk} \log \det(H(k) - E) \qquad (3)$$

Clearly, the phase diagram shown in Fig. 1(b) exhibits symmetry with respect to the Hermitian line $t_3 = t_4$, as the orientation of NHSE is determined by the relative magnitudes between $t_3$ and $t_4$. Specifically, when $t_3 > t_4$ ($t_3 < t_4$), all OBC eigenmodes are located towards the left (right) boundary. Interestingly, NHSEs arise solely from non-reciprocal couplings perpendicular to the direction of wave propagation. Furthermore, tuning the ratio $t_2/t_1$ allows for further control over the phase diagram of this model. When $t_2 < t_1$, there is a reversal in the direction of NHSE within certain phase regions.

Finally, we emphasize the significance of *GT* symmetry in elucidating the distinctive characteristics of both energy bands and the GBZ. We define the operator $\Theta$ as $\Theta = GT$, where *GT* represents the combined glide and time-reversal symmetries. The action of $\Theta$ on the state $\psi_{nk}$ is given by $\Theta^2 \psi_{nk} = e^{ik} \psi_{nk}$, where $n$ denotes the band index and $k$ signifies the wave vector. In the Hermitian case, when $k = \pm\pi$, we have $\Theta^2 = -1$. In the non-Hermitian scenario, with $k = \pm\pi + i\gamma$ ($\gamma \in \mathbb{R}$), we have $\Theta^2 = -e^{-\gamma}$. These results lead to a Kramers-like double degeneracy, meaning that both the real and imaginary parts of the band structure inevitably exhibit degeneracy at the Brillouin zone boundaries. More specifically, considering the chiral symmetry of this model, both the real and imaginary parts of the band structure have two doubly degenerate points at the Brillouin zone boundaries. Similarly, GBZ also possesses such characteristics, which will be elaborated in the next section.

## III. PHASE DIAGRAM ANALYSIS UNDER OBC

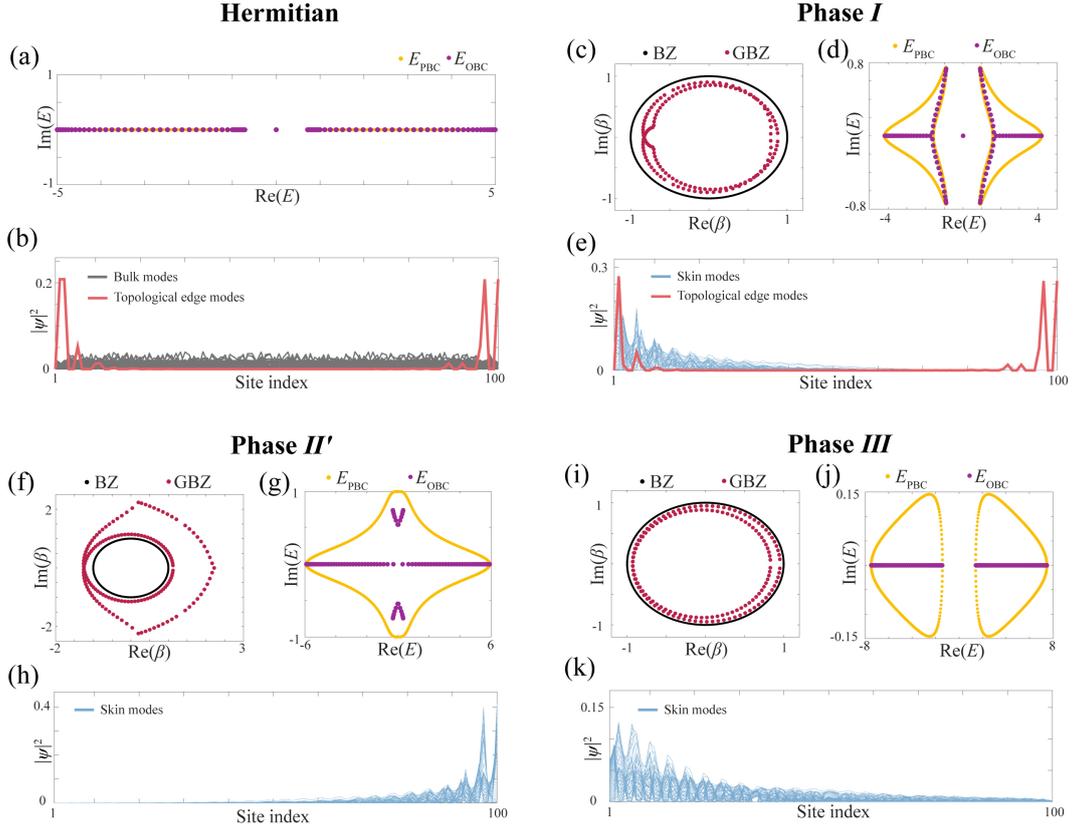

FIG. 2. (a), (d), (g), and (j) The PBC and OBC complex energy spectra represented by yellow and purple dots, respectively. (c), (f), and (i) The first Brillouin zones(BZ) and GBZs indicated by black and red dots, respectively. The GBZs are calculated with 160 sites. (b), (e), (h), and (k) exhibit all OBC eigenmodes, where the red lines represent topological edge modes, brown lines denote bulk modes, and blue lines indicate skin modes. The four selected parameter sets $(t_3, t_4)$: Hermitian(2,2), phase $I$ (2,0.5), phase $II'$(2,4), phase $III$(5,4).

In this section, we will investigate the band topology and non-Hermitian characteristics of this $GT$ model by employing complex wavevector spectra, complex energy spectra, and eigenmodes. To facilitate the systematic investigation of eigenmode properties across different phase regions, we have selected three aforementioned parameters sets and a Hermitian case $t_3 = t_4 = 2$.

Notably, both the Hermitian case and Phase *I* case exhibit topologically non-triviality, indicating the presence of topological edge modes. The manifestation of zero-energy states within the gap is evident in their OBC complex energy spectra, as depicted in Figs. 2(a) and (d). These zero-energy eigenstates correspond to topological edge modes, which are visually emphasized by the red lines in Figs. 2(b) and (e). However, their bulk modes exhibit significant differences due to non-reciprocity. In the Hermitian case, the bulk modes extend uniformly throughout the entire bulk region. In contrast, all bulk modes in Phase *I* tend to localize towards left boundaries, which exemplifying the presence of NHSE. To elucidate the origin of NHSE, we analyze both the GBZ and the winding numbers derived from the PBC complex energy spectra. In the view of the GBZ, each SSH chain in the non-Hermitian *GT* model has its own GBZ. When two chains are coupled with non-reciprocal interchain coupling, two GBZs combine and form the associated respective GBZ of this two-chain model[46], as shown in Figs. 2(c), (f), and (i). On one hand, all complex wave vectors $\beta$ within the GBZ possess magnitudes less than 1, indicating that all skin modes localize towards the left boundary. On the other hand, the PBC complex energy spectra form two loops with non-zero enclosed area which further confirms NHSE's existence. More specifically, the winding number $W_E$ of the complex energy spectra is $-1$, indicating a leftward direction for NHSE. The aforementioned evidence collectively validates the presence and direction of NHSE.

The Phase *II'* case, as depicted in Figs. 2(f)-(h), is characterized by its gapless nature, thereby precluding the existence of topological edge states. Moreover, due to $t_3 < t_4$, all skin modes exhibit a rightward orientation. This observation is further substantiated by the fact that all $\beta$ values within the GBZ possess magnitudes greater than 1 and the winding number of the PBC complex energy spectra equals 1. Similarly, for the Phase *III* case illustrated in Figs. 2(i)-(k), it exhibits topological triviality with no emergence of topological edge states within its gap region. Additionally, with $t_3 > t_4$, NHSE directionality is observed towards the left side as confirmed both by examining the GBZ and analyzing the winding number of the PBC complex energy spectra. Especially, for all three non-Hermitian cases considered herein, their respective GBZs consist of two loops intersecting at a point where $\text{Re}(\beta) < 0$ and $\text{Im}(\beta) = 0$, a direct consequence of *GT* symmetry as previously mentioned. At this intersection point, $\beta$ aligns with the boundary of the BZ where $\text{Re}(k) = \pi$.

## IV. DYNAMIC PHASE AND CLASSIFICATION OF THE NON-HERMITIAN *GT* MODEL

In this section, we will demonstrate the rich dynamic behavior of the non-Hermitian *GT* model, and also presenting a method to discern various dynamic phases. This methodology will serve as an indispensable guide for subsequent analysis of the system's energy amplification characteristics.

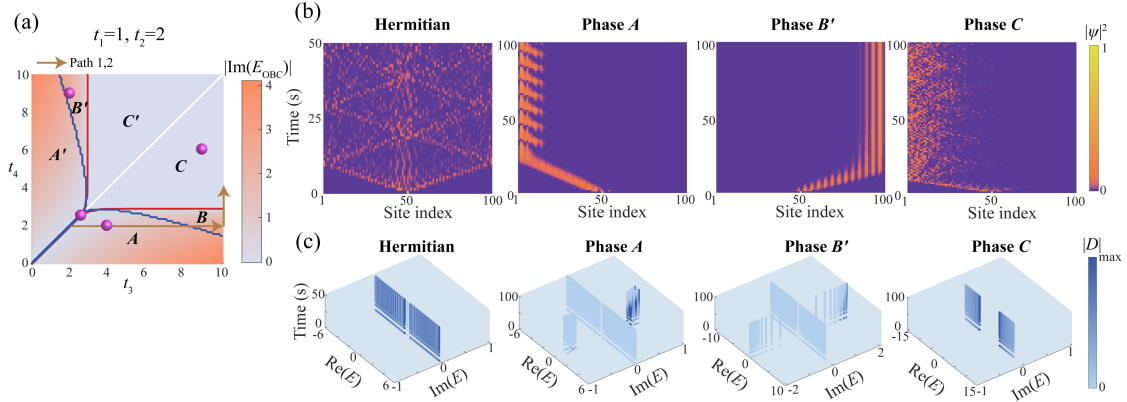

FIG. 3. Dynamic phase diagram of the non-Hermitian *GT* model and distinct dynamic behaviors. (a) Dynamic phase diagram. The white line represents the Hermitian condition, indicating $t_3 = t_4$. Red curves signify the imaginary gap-closing boundary, where $\text{Im}(E_{\text{OBC}}) = 0$. Blue curves denote the phase boundary between phases *A* (*A′*) and *B* (*B′*), used to identify whether the dominant frequency in long-time evolution is double or singular. The brown arrows represent two paths 1 and 2 examining the time-evolution of the system's energy. The purple spheres represent the four selected $(t_3, t_4)$ parameter sets: Hermitian (2.5,2.5), phase *A* (4,2), phase *B′* (2,9), and phase *C* (9,6). (b) The spatiotemporal profiles $|\psi(t)|^2$ for four cases are normalized at each instant. (c) $|D(t)|$ is derived by decomposing the time-dependent wavefunction $\psi(t)$ for four cases into OBC eigenmodes and is normalized at each instant.

## A. Dynamic phase diagram of the non-Hermitian *GT* model and OBC decomposition

The eigenvalues of non-Hermitian systems, which can potentially be complex due to their inherent non-Hermitian nature, often possess an imaginary component that directly corresponds to system's energy amplification or attenuation. Consequently, the temporal evolution of non-Hermitian systems exhibits substantial deviations from that of Hermitian systems.

As the central focus of this section, we present a comprehensive theoretical derivation to elucidate the distinct dynamic phases. We propose that a crucial aspect in understanding non-Hermitian dynamics lies in identifying the dominant eigenmodes governing the temporal evolution. Consequently, we decompose the time-dependent wavefunction $\psi(t)$ into the OBC eigenmodes.

At any given moment, the final state $\psi(t)$ can be obtained by applying the time evolution operator $U \equiv e^{-iH_{OBC}t}$ to the initial state $\psi(0)$.

$$\psi(t) = e^{-iH_{OBC}t} \cdot \psi(0). \tag{4}$$

The initial state $\psi(0)$ is excited by the delta function $\delta_{x,x_0}$, where $x_0$ corresponds to the middle site of $H_{OBC}$. A delta function $\delta_{x,x_0}$ encompasses a broad spectrum of frequencies, thereby unveiling comprehensive dynamic properties. Then, we can decompose the final state $\psi(t)$ by the OBC eigenmodes

$$\psi(t) = \sum_j D_j(t) |\varphi_{j,R}\rangle, \tag{5}$$

where $j$ is the degree of freedom of OBC Hamiltonian, $|\varphi_{j,R}\rangle$ and $\langle\varphi_{j,L}|$ are the right and left eigenvectors of this OBC Hamiltonian, respectively. The right and left eigenvectors can be obtained using the following equation.

$$H_{OBC}|\varphi_{j,R}\rangle = E_j|\varphi_{j,R}\rangle, \tag{6}$$

$$\langle\varphi_{j,L}|H_{OBC} = E_j\langle\varphi_{j,L}|. \tag{7}$$

Additionally, $\varphi_{j,R}$ and $\varphi_{j,L}$ satisfy the biorthogonality condition,

$$\langle\varphi_{i,L}|\varphi_{j,R}\rangle = \delta_{i,j}. \tag{8}$$

This yields the weight coefficients $D_j(t)$, which represents the contribution of the $j$-th eigenmodes and is generally complex:

$$D_j(t) = \langle\varphi_{j,L}|\psi(t). \tag{9}$$

Based on the distribution of $D_j(t)$, we can construct a comprehensive dynamic phase diagram for the non-Hermitian *GT* model, as depicted in Fig. 3a. Notably, the dynamic phases and eigenmode phases exhibit remarkable differentiation. The sole similarity lies in the fact that the orientation of dynamic NHSE also demonstrates symmetry with respect to the Hermitian line $t_3 = t_4$, indicating that when $t_3 > t_4(t_3 < t_4)$, dynamic NHSE is directed towards the left (right) boundary.

First, let us commence by analyzing the dynamics of Hermitian systems. It is evident that in the case of Hermitian scenario depicted in Fig.3(c), all eigenmodes, except for the edge states, exhibit significant magnitudes of $|D(t)|$ over time. This can be attributed to the fact that the discrete Fourier transform of the delta function from coordinate space to momentum space yields a constant value, representing excitation across all momenta $k$. Additionally, the OBC energy remains purely real, signifying the absence of any amplification or attenuation in individual states. Consequently, all bulk states dominate the dynamics, with the edge states being excluded. Furthermore, the evolutive state in this Hermitian scenario in Fig. 3(b) demonstrates conventional dynamic behavior characteristic of Hermitian systems.

Notably, these wave packets tend to propagate leftward or rightward or remain stationary and undergo elastic scattering upon encountering boundaries.

Furthermore, we present a comprehensive analysis of phases $A$ and $B'$ together (a similar analysis applies to phases $A'$ and $B$). The discrepancy between these two phases arises from the disparity in dominant frequencies during long-term evolution: phase $A$ exhibits dual dominant frequencies, whileas phase $B'$ only possesses one. In both phases $A$ and $B'$, the OBC eigenvalues manifest as complex numbers, with their imaginary part indicating whether these states undergo amplification or attenuation over time. Consequently, states with the largest imaginary parts will experience significant amplification, ultimately exerting dominance over the dynamics. To gain a more intuitive understanding, we employ Green's function. The discrete Fourier transform of the matrix element of Green's functions [38] in the time domain is expressed as:

$$G_{ij}(t) = \int \frac{d\omega}{2\pi} G_{ij}(\omega) e^{-i\omega t}, \tag{10}$$

where $\omega$ is the complex frequency, $j$ is the site of the input source with the $\omega$-frequency, $i$ is the site of the measured signal response. The magnitude of $e^{-i\omega t}$ increases with higher imaginary parts of $\omega$, indicating that the integral elements of the complex frequency with maximum imaginary parts contribute significantly during time evolution. Additionally, both phases $A$ and $B'$ wave functions demonstrate a tendency for the wave packet to move towards one of boundaries (left boundary for phase $A$ and right boundary for phase $B'$) and localize at the boundary upon reaching it, suggesting dynamic NHSE. Interestingly, in phase $A$, the wave functions exhibit a pronounced beating phenomenon due to the simultaneous effect of two frequencies, providing evidence for dual-frequency dominance in this phase. In contrast, phase $B'$ does not display a beating phenomenon as it is characterized by a single dominant frequency.

Finally, our attention is directed towards phase $C$ (with a similar counterpart, phase $C'$), which not only exhibits dynamic NHSE but also showcases purely real eigenvalues. Consequently, this phase demonstrates a scenario akin to Hermitian cases where all bulk states significantly contribute to the temporal evolution, as depicted in Fig. 3(c). Notably, the evolution converges towards the modes located near the line-gap edges, which correspond to the modes with the smallest $|\beta|$ (this will be further analyzed in subsequent sections). From a wavefunction perspective, an abundance of purely real eigenvalues implies that numerous modes are involved in the dynamics, leading to complex interference that disrupts any potential beating phenomenon. Therefore, the wave functions will exhibit dynamic NHSE and appear relatively chaotic and complex behavior. Moreover, it is observed that certain wave packets display exit velocities inconsistent with their incident velocities, indicating the occurrence of inelastic scattering.

## B. Understanding dynamic amplification through OBC eigenmodes

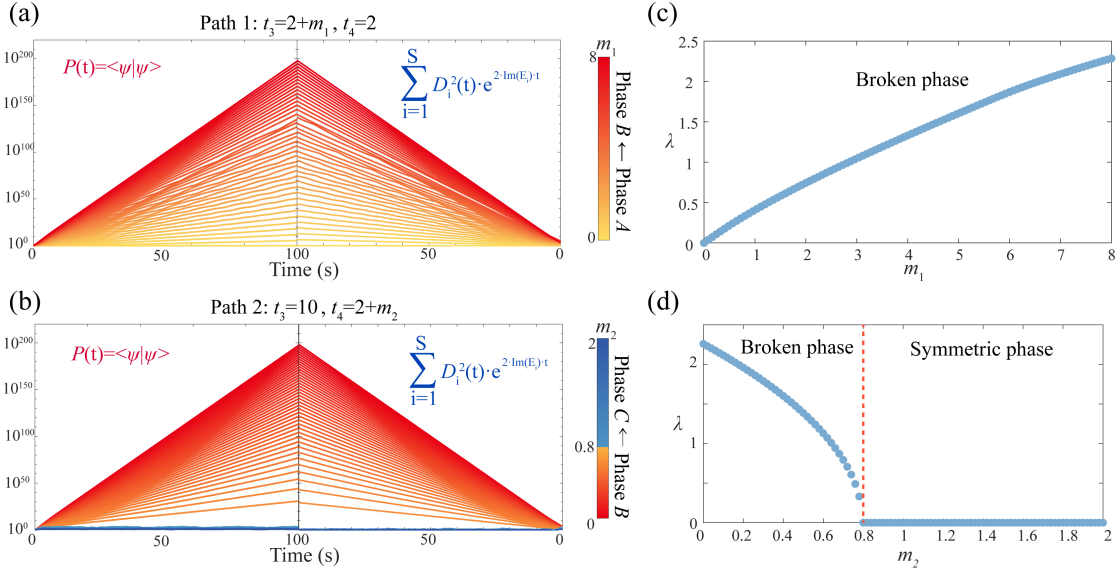

FIG. 4. Understanding dynamic amplification through OBC eigenmodes. (a)-(b) The time-evolution of the system's energy for two paths in Fig. 2(a). For each subplot, the left depicts the system's energy computed from the final state, while the right illustrates the system's energy obtained through the weighted summation of individual energies. (c)-(d) Lyapunov exponent $\lambda$ along paths 1 and 2. Path 1: $t_3 = 2 + m_1$, $t_4 = 2$. Path 2: $t_3 = 10$, $t_4 = 2 + m_2$. $m_1$ and $m_2$ are indicated by the color bar. $t_1 = 1$ and $t_2 = 2$.

In the previous calculations, we normalized the wave functions. However, in general non-Hermitian dynamics, the conservation of system's energy (or the norm of the wave function) is generally not observed. The imaginary part of the OBC energy plays a crucial role in energy amplification. Here, we will conduct both qualitative and quantitative analyses to investigate the amplification of system's energy during time evolution, enabling us to identify phase transitions in non-Hermitian dynamics.

Primarily, we will perform derivations to establish the relationship between the time-dependent wave function and the imaginary component of energy. It is well-known that the time evolution operator can also be represented as an expansion in terms of eigenstates:

$$U \equiv \sum_j |\varphi_{j,R}\rangle e^{-iE_j t} \langle \varphi_{j,L}|. \tag{11}$$

We attain the final state again

$$\psi(t) = \sum_j |\varphi_{j,R}\rangle e^{-iE_j t} \langle\varphi_{j,L}|\psi(0). \tag{12}$$

Separating the real part $\text{Re}(E_j)$ and the imaginary part $\text{Im}(E_j)$ of the complex energy $E_j$, we have

$$\psi(t) = \sum_j |\varphi_{j,R}\rangle e^{\text{Im}(E_j)\cdot t} e^{-i\cdot \text{Re}(E_j)\cdot t} \langle\varphi_{j,L}|\psi(0). \tag{13}$$

Substituting it into Eq. (9), we obtain:

$$D_j(t) = \langle\varphi_{j,L}| \sum_i |\varphi_{i,R}\rangle e^{\text{Im}(E_i)\cdot t} e^{-i\cdot \text{Re}(E_i)\cdot t} \langle\varphi_{i,L}|\psi(0). \tag{14}$$

To avoid numbering conflicts, let us re-index the degrees of freedom in Eq. (14) as "$i$". Due to the biorthogonality property in Eq. (8), we have $D_j(t) \neq 0$ only when $i = j$, thus yielding:

$$D_j(t) = e^{\text{Im}(E_j)\cdot t} e^{-i\cdot \text{Re}(E_j)\cdot t} \langle\varphi_{j,L}|\psi(0). \tag{15}$$

Taking into account that the final state $\psi(t)$ in the previous analysis is normalized, $D_j(t)$ actually do not include the component $e^{\text{Im}(E_j)\cdot t}$. In other words,

$$D_j(t) = e^{-i\cdot \text{Re}(E_j)\cdot t} \langle\varphi_{j,L}|\psi(0). \tag{16}$$

Consequently, we can conclude that $D_j(t)$ effectively represents the weight of $e^{\text{Im}(E_j)\cdot t}$. By weighting $e^{\text{Im}(E_j)\cdot t}$ with $D_j(t)$ and combining with Eq. (5), the probability $\langle\psi(t)|\psi(t)\rangle$ can be obtained as

$$\langle\psi(t)|\psi(t)\rangle = \sum_j \langle\varphi_{j,L}| e^{2t\cdot \text{Im}(E_j)} \cdot D_j(t)^2 |\varphi_{j,R}\rangle. \tag{17}$$

By utilizing the biorthogonality property in Eq. (8), the above equation can be simplified to

$$\langle\psi(t)|\psi(t)\rangle = \sum_{j=1}^{S} e^{2t\cdot \text{Im}(E_j)} \cdot D_j(t)^2, \tag{18}$$

where $S$ is the number of sites.

We calculate both sides of the Eq. (18) for two paths with the corresponding results presented in Figs. 4(a)-(b). The wave amplification values, determined by the interplay between $D_j(t)$ and $\text{Im}(E_j)$, exhibit consistency with the probability for both paths. Notably, smooth behavior is observed in the wave amplification values along path 1 due to the unconserved phase present in both phases $A$ and $B$. In contrast, a sudden transition is observed in the wave amplification values along path 2 near the boundaries between phases $B$ and $C$, attributed to the presence of a conserved phase in phase $C$. This signifies a transition from the non-Bloch *PT* broken phase to the non-Bloch *PT* symmetric phase.

We further emphasize that the non-Bloch phase transition can be effectively characterized by the "not smooth" behavior observed in the Lyapunov exponent $\lambda(v)$[47] at zero drift velocity $v$, where $\lambda(v) = \text{Im}(E(k_s)) - v\text{Im}(k_s)$, with $v = (\frac{dE}{dk})_{k_s}$ representing the drift velocity and $k_s$ denoting the dominant saddle point. The Lyapunov exponents along two distinct paths are illustrated in Figs. 4(c)-(d). These findings unveil the occurrence of non-Bloch symmetric-broken phase transitions along path 2, which aligns with the behavior of $P(t)$. Another convenient approach to ascertain the occurrence of a non-Bloch *PT* phase transition is through the observation of the GBZs. When the system undergoes a transition from non-Bloch *PT* symmetric phase to the non-Bloch *PT* broken phase, saddle points in the GBZ shift from existence to absence. This phenomenon can also be discerned in Figs. 2(c), (f), and (i).

## V. NON-HERMITIAN DYNAMICS ANALYSIS BY Z TRANSFORMATIONS

### A. Z transformation to the GBZ

In Hermitian theory of crystals, the dynamics are governed by the dispersion relation, which describes the relationship between wave vector (or momentum) $k \in \mathbb{R}$ and $\omega \in \mathbb{R}$. However, in non-Hermitian systems, Bloch theory is extended to non-Bloch theory where $k$ is reformulated as a complex momentum, or the generalized wave vector $\beta \equiv e^{iq}$, $q \in \mathbb{C}$, to account for the NHSE. This implies that the BZ, which is a unit circle $e^{ik}$ with $k \in [-\pi, \pi]$ in the complex plane, transforms into a GBZ, which is typically not circular. Similar to Hermitian band theory, we analyze the dynamics of our non-Hermitian system by projecting the evolving wave function in coordinate space onto the GBZ through Z transformations and decomposing it into PBC eigenmodes as detaily described in this section. We can calculate time-dependent components $|G|$ to gain insights into states with predominant contributions during dynamic evolution.

Here, we present a theoretical demonstration of how to obtain the time-dependent components $|G|$ derived through Z transformations. Specifically, we expand from the wave vector $k$ in the BZ to the generalized wave vector $\beta$ in the GBZ by utilizing the identity $\beta \equiv e^{iq}$, where $q \in \mathbb{C}$ represents a usual wave vector. Thus, for a given 1D tight-binding Hamiltonian $H(k)$, one obtains an analytically continued Hamiltonian $H(\beta)$:

$$H(k) \to H(\beta). \tag{19}$$

Similarly, the OBC final state $\psi(t,n)$ in the coordinate space at any given time can be expressed

$$|\psi(t,n)\rangle = U(t,t_0)|\psi(t_0,n)\rangle, \tag{20}$$

where $n$ denotes the site index, $U(t,t_0)$ still represents the time evolution operator $U(t,t_0) = e^{-iH_{OBC}t}$ and an initial state $\psi(t_0,n)$ is excited by the delta function $\delta_{x,x_0}$ on a Hamiltonian with OBC. Subsequently, we can perform a Z transformation for $\psi(t,n)$ to acquire its representation $\Psi_j(t,\beta)$ in the generalized momentum space

$$\Psi_j(t,\beta) = \sum_{n=1}^{N} \psi(t,n)\beta^{-x_n}, \tag{21}$$

where $x_n$ represents the $x$ coordinate value of the $n$th site. Through the analysis of the final state $\Psi_j(t,\beta)$ in the generalized momentum space, one can identify the generalized wave vectors $\beta$ that make significant contributions during the evolution process. Subsequently, we decompose the final state $\Psi_j(t,\beta)$ into the PBC eigenmodes of the non-Bloch Hamiltonian and determine their corresponding weight coefficients $G_j(t,\beta)$,

$$|\Psi_j(t,\beta)\rangle = \sum_j G_j(t,\beta)|\varphi_{j,R}(\beta)\rangle, \tag{22}$$

$$G_j(t,\beta) = \langle\varphi_{j,L}(\beta)|\Psi_j(t,\beta)\rangle, \tag{23}$$

where $\varphi_{j,R}(\beta)$ and $\varphi_{j,L}(\beta)$ are the right and left eigenvectors of the non-Bloch Hamiltonian, respectively, and can be obtained using Eqs. (6) and (7).

It is crucial to highlight that the setting of the zero point position in a one-dimensional chain significantly influences the outcomes of Z transformation analysis. The impact of a translation operation on $x_0$ can be mathematically described within the framework of Z-transform, considering a coordinate sequence $\xi(x)$. The original Z-transform is:

$$Z\{\xi(x)\} = \sum_{x=-\infty}^{\infty} \xi(x)\beta^{-x}. \tag{24}$$

After shifting the sequence by $x_0$, the Z-transform becomes

$$Z\{\xi(x-x_0)\} = \sum_{x=-\infty}^{\infty} \xi(x-x_0)\beta^{-x}. \tag{25}$$

We set $u = x - x_0$ to simplify this expression.

$$Z\{\xi(x-x_0)\} = \sum_{u=-\infty}^{\infty} \xi(u)\beta^{-(u+x_0)} = \beta^{-x_0}\sum_{u=-\infty}^{\infty}\xi(u)\beta^{-u}. \tag{26}$$

Therefore, we have

$$Z\{\xi(x-x_0)\} = \beta^{-x_0}Z\{\xi(x)\}. \tag{27}$$

Consider shifting the entire sequence to the right by half of its length, denoted as $x_h$, in order to place the origin at the leftmost end of the sequence. In this case, $x_h$ is greater than 0 and the function $\beta^{-x_h}$ exhibits a monotonically decreasing trend with respect to $\beta$. Consequently, smaller values of $\beta$ will yield larger coefficients for $\beta^{-x_h}$ after undergoing a Z-transform. This implies that compared to placing the origin in the middle of the chain, smaller values of $\beta$ will experience a greater enhancement in coefficient magnitude following a Z-transform. This phenomenon presents a dual nature, with potential advantages and disadvantages. On one hand, strategically placing the zero point in unidirectional NHSE can effectively accentuate the significant contributions of $\beta$. Conversely, it may introduce distortions to the transformation outcomes. Therefore, we recommend selecting zero point coordinates based on the specific focus of analysis when employing this methodology. It is crucial to note that in systems featuring bipolar NHSE, positioning the zero point at its midpoint becomes imperative to prevent an excessive emphasis on one side's NHSE contribution.

Similarly, the evolving wave function in coordinate space can be projected onto the BZ using discrete Fourier transformations, allowing for decomposition into PBC eigenmodes and calculation of the weighting coefficients $|K|$. Additionally, by incorporating energy bands, dynamic behavior analysis of non-Hermitian systems becomes feasible.

**B. Dynamics phase diagram analysis**

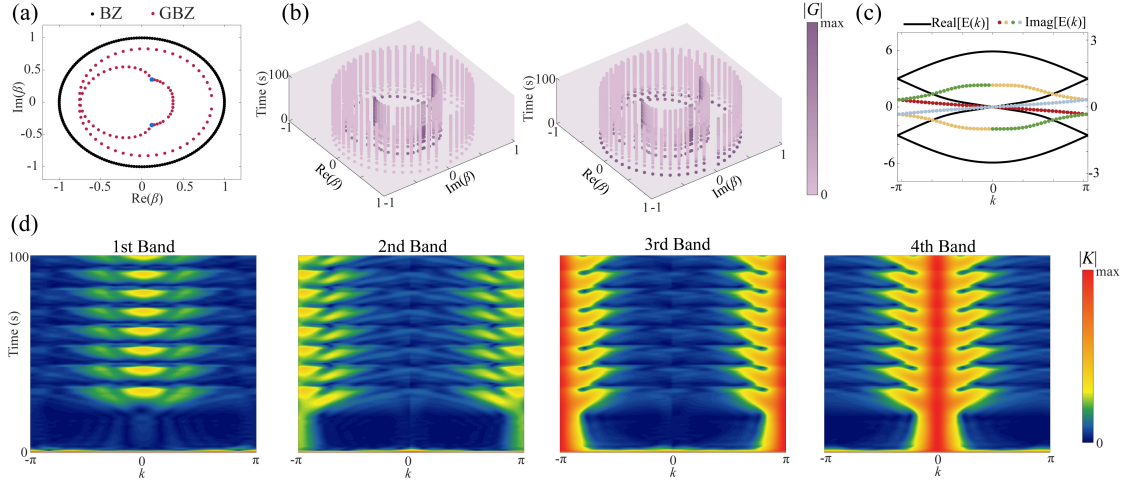

FIG. 5. (a) The GBZ for phase $A$. Black dots represent the first BZ and red dots indicate the GBZ. Blue dots are enlarged to emphasize the primary generalized wave vector $\beta$ during long-time evolution. (b) $|G|$ is derived by Z transformations on the time-dependent wavefunction $\psi(t)$ and is normalized at each instant. The Time-axis is represented on a logarithmic scale. The results are mapped to the corresponding GBZ. $|G|$ with negative eigenvalues on the left side and $|G|$ with positive eigenvalues on the right side. (c) Energy bands $E(k)$ for phase $A$. The real parts of $E(k)$ are denoted by black curves and associated with the left scale axis. The imaginary parts of four energy bands are marked in red, orange, green and dusty blue, respectively, and associated with the right scale axis. (d) $|K|$ for the four bands as a function of $k$ and $t$. The results are mapped to the corresponding BZ. The parameters: $t_1 = 1$, $t_2 = 2$, $t_3 = 4$ and $t_4 = 2$.

In this section, we will analyze each case individually, starting with phase $A$. Additionally, the $GT$ symmetry also gives rise to a complex geometry for the GBZ. In general non-Hermitian systems, the GBZ is a closed curve in the complex wavevector plane around the origin, typically deviating from a circular shape. According to the previous discussion, GBZ will exhibit a Kramers-like double degeneracy and intersect at a point where $\text{Re}(\beta) < 0$ and $\text{Im}(\beta) = 0$. Upon substituting each $\beta$ into the non-Bloch Hamiltonian, four eigenvectors are obtained; however, only two of them correspond to the eigenvalues of $H_{obc}$. We exclusively apply the decomposition to these two eigenvectors, resulting in $|G|$ having two components. Considering that the eigenvalues of these two eigenvectors satisfy $E$ and $-E$ respectively, we present $|G|$ with negative eigenvalues on the left side of Fig. 5(b) and $|G|$ with positive eigenvalues on the right side. Consequently, Fig. 5(b) displays two $G_j$ components accordingly. At time $t = 0$, all $\beta$ are excited due to the real-space delta function $\delta_{x,x_0 = middle\ site}$. It is noteworthy that the wavefunction evolves towards eigenmodes characterized by the smallest $|\beta|$. The integral formula based on GBZ for calculating matrix elements of Green's functions [38] can be expressed as follows:

$$G_{ij}(\omega) = \int_{GBZ} \frac{\partial \beta}{2\pi i \beta} \frac{\beta^{i-j}}{\omega - H(\beta)}. \tag{28}$$

When $|G_{ij}(\omega)| \gg 1$, it indicates that the propagation of signals is amplified. Conversely, when $|G_{ij}(\omega)| \ll 1$, the signal is attenuated during propagation. In phase A, our emphasis is directed towards

$$G_{1,x_0=middle\ site}(\omega) = K \int_{GBZ} \frac{\partial \beta}{2\pi i \beta} \frac{\beta^{-(x_0-1)}}{\omega - H(\beta)}, \tag{29}$$

where $K$ is a factor of order unity, $x_0$ is the middle site of OBC chains. Due to the non-Hermitian dynamics, the condition $\left|G_{1,\frac{NU}{2}}(\omega)\right| \gg 1$ ensures a dominant contribution from the smallest $|\beta|$ in this integral equation. The eigenmodes associated with these $|\beta|$ values are localized specifically at the left boundary. Consequently, during the evolutionary process, wave packets exhibit a pronounced tendency to move towards the left direction due to dynamic NHSE.

Similarly, by applying discrete Fourier transform to project the evolved state onto the first BZ and then decomposing it onto the eigenvectors of the Bloch Hamiltonian, we can extract wavevector information to comprehend dynamic NHSE. The weight components $|K|$ for the four bands can be calculated and is normalized at each instant, displayed in Fig. 5(d). At time $t = 0$, all wave vectors $k$ have been also excited because of the real-space delta function $\delta_{x,x_0=middle\ site}$. The highest component for the first and fourth bands is observed at $k = 0$, while for the second and third bands it occurs at $k = \pm\pi$. The scenarios involving high component $k$ vary between the wave packet's movement process and its localized state after reaching the boundary. Throughout the movement process of the wave packet, these high-components of $k$ remain relatively stable.

However, following localization at the boundary, the distribution of high-components exhibit periodicity, which is in accordance with the oscillation period of the wave packet at the boundary in real-space wave function. This phenomenon can also be attributed to two dominate frequencies. By further considering properties of energy bands depicted in Fig. 5(c), a comprehensive understanding of dynamic NHSE can be achieved. We define the real parts of energy bands as the partial derivative of energy $Re(E)$ with respect to the wavevector $k$, commonly known as the group velocity, denoted as

$$v_{group} = \frac{\partial Re[E(k)]}{\partial k}. \tag{30}$$

Additionally, the imaginary part of energy eigenvalues can indicate either amplification or attenuation during temporal evolution. Initially, when considering the wave vectors at $k = 0$ for 1st and 4th bands, both the group velocity $v_{group}$ and the imaginary part of the energy eigenvalues converge to zero. This convergence indicates that wave packets are inclined to remain stationary without experiencing amplification or attenuation during time evolution. Subsequently, upon examining the wave vectors at $k = -\pi$ for the second band and $k = +\pi$ for the third band, a positive group velocity emerges alongside negative imaginary part of energy eigenvalues. This configuration implies a tendency for wave packets to propagate towards the right direction while experiencing attenuation over time. Conversely, when the wave vectors are situated at $k = +\pi$ for the second band and $k = -\pi$ for the third band, the group velocity becomes negative, accompanied by a positive imaginary part of the energy eigenvalues. This scenario suggests a propensity for wave packets to move towards the left direction and undergo amplification during evolutionary process. In summary, wave packets exhibiting an inclination towards amplification will manifest most prominently, consequently driving the dynamic NHSE towards the left boundary.

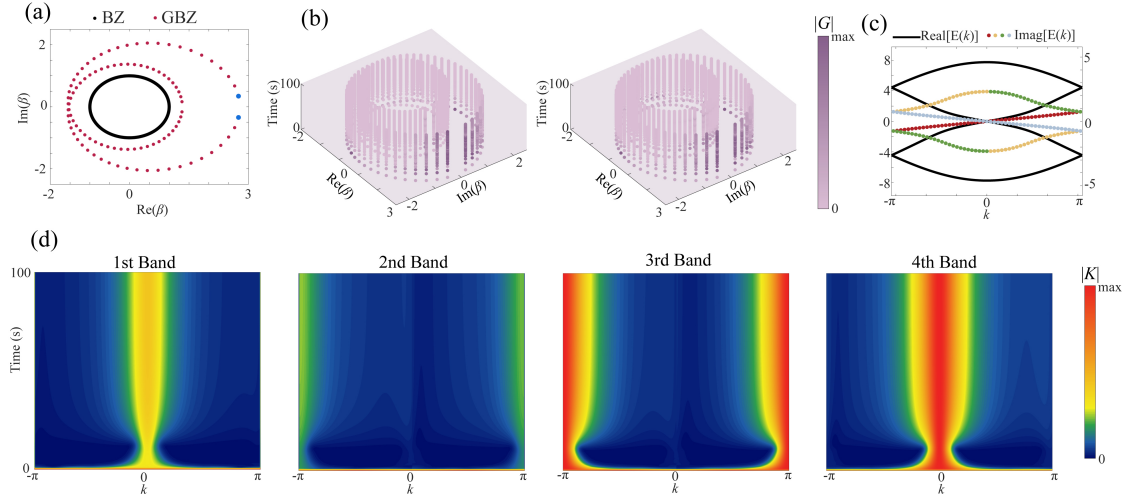

FIG. 6. Dynamics analysis for phase $B'$. (a) The GBZ. (b) The calculation results of applying the Z transformation to GBZ. (c) Energy bands $E(k)$. (d) The calculation results of applying the Fourier transformations to BZ. The parameters: $t_1 = 1$, $t_2 = 2$, $t_3 = 2$ and $t_4 = 9$.

In the case of phase $B'$, due to the reversed relationship between $t_3$ and $t_4$, there is a shift of dynamic NHSE towards the right direction. Simultaneously, all $|\beta|$ in the GBZ will exceed unity, as depicted in Fig. 6(a), indicating that all eigenmodes are localized on the right boundary. In this scenario, the integral form of matrix elements of the Green's function takes on the following form:

$$G_{x_r,x_0=middle\ site}(\omega) = K \int_{GBZ} \frac{\partial \beta}{2\pi i \beta} \frac{\beta^{-(x_0-x_r)}}{\omega - H(\beta)}, \tag{31}$$

where $x_r$ is the far-right site of OBC chains. The positive exponential factor of $\beta$ can be observed, indicating its increasing contribution in the integration. Consequently, $\beta$ with the maximum magnitude holds the highest component, aligning with the findings depicted in Fig. 6(b). These dominant $\beta$ values with maximum magnitudes govern the rightward dynamic NHSE.

Another noteworthy observation is that the high component $|K|$ positions in the first BZ remain unchanged (as shown in Fig. 6(d)). Specifically, they are still located at $k = 0$ for the 1st and 4th bands, and at $k = \pm\pi$ for the 2nd and 3rd bands. However, there is a modification in the characteristics of energy bands (refer to Fig. 6(c)). For energy exhibiting positive group velocity, its imaginary part exceeds zero; conversely, for energy displaying negative group velocity, its imaginary part falls below zero. Consequently, the phenomenon of dynamic NHSE stands in stark contrast to that observed within region $O$. During time evolution, wave packets moving towards the right gradually amplify themselves, thereby manifesting dynamic NHSE in a rightward direction. Furthermore, owing to phase $B'$ being predominantly single-frequency dominant compared to phase $A$, it does not exhibit periodic oscillations akin to those observed in phase $A$.

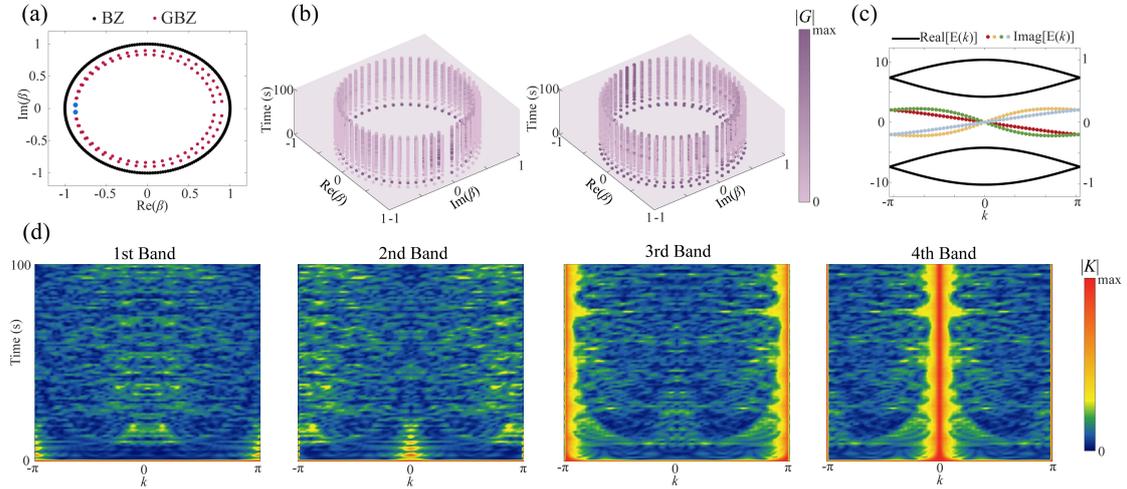

FIG. 7. Dynamics analysis for phase $C$. (a) The GBZ. (b) The calculation results of applying the Z transformation to GBZ. (c) Energy bands $E(k)$. (d) The calculation results of applying the Fourier transformations to BZ. The parameters: $t_1 = 1$, $t_2 = 2$, $t_3 = 9$ and $t_4 = 6$.

For phase $C$, it is similar to phase $A$ and can be analyzed using the same method combined with Fig.7 to obtain all the results. Interestingly, it can be observed that the magnitudes of components $|K|$ exhibit a higher degree of chaos in comparison to phases $A$ or $B'$, which can be attributed to the purely real OBC energy spectra of phase $C$ or $C'$. As a result, all eigenmodes collectively exert an influence on the dynamic evolution, leading to the emergence of a relatively chaotic state that is observable in the spatiotemporal profile.

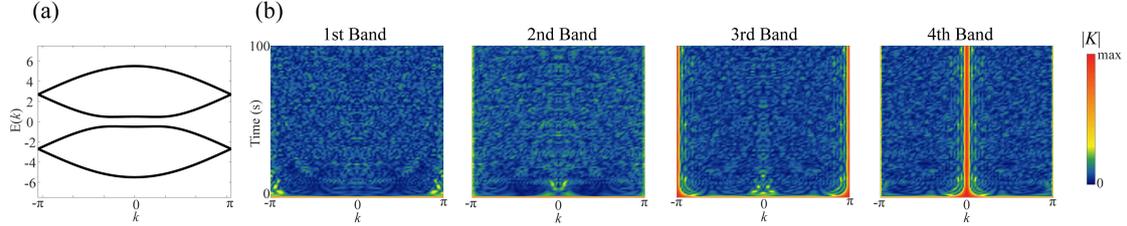

FIG. 8. Dynamics analysis for the Hermitian case. (a) Energy bands $E(k)$. (b) The calculation results of applying the Fourier transformations to BZ. The parameters: $t_1 = 1$, $t_2 = 2$, $t_3 = t_4 = 2.5$.

In the Hermitian limit, the generalized wave vector $\beta$ collapses to an purely real wave vector $k$, and Z transformation degenerates into a discrete Fourier transformation. As a result, decomposition analysis of the GBZ and first BZ yields identical results. Therefore, this discussion primarily focuses on the first BZ. Upon observing Fig. 8(b), it is evident that the 3rd band at $k = \pm\pi$ and the 4th band at $k = 0$ exhibit relatively significant components. However, in Hermitian systems depicted in Fig. 8(a), the bands exhibit purely real behavior without any amplification or attenuation. The collective influence of high wavevector components induces wave packets to demonstrate tendencies towards leftward movement, rightward movement, or stationary behavior, while also experiencing elastic scattering upon encountering boundaries. This distinction highlights a notable disparity from non-Hermitian systems.

## VI. Conclusion and discussions

In this work, we theoretically study a 1D non-Hermitian model based on a double-chain system with $GT$ symmetry. This model has a number of intriguing properties in both periodic and finite systems, showing unconventional eigenstates and dispersions. In most parameter regions, the model has NHSEs of which the direction depends on the non-reciprocal interchain couplings. We analyze the properties of the eigenstates and dispersions carefully and establish the phase diagram for the periodic systems. Besides, the model showcases rich dynamic phenomena, including dynamic NHSEs in various directions, amplification of the system's energy. Moreover, the wave dynamics at the boundaries can be oscillatory, non-oscillatory, or chaotic, while the wave amplitude can be oscillatory, exponentially increasing, or decreasing. By decomposing the time-dependent wave function into the OBC eigenmodes, we gain insight into the dynamics of non-Hermitian systems, recognizing that the dominant contribution arises from eigenmodes exhibiting the largest imaginary energy. This decomposion allows us to connect the dynamic properties with the phase diagram for finite systems. Furthermore, by transforming the time-dependent wave function onto the GBZ and subsequently decomposing it into non-Bloch eigenmodes, we elucidate the inherent relationship between the dominant generalized wavevector $\beta$ in dynamics and the direction of dynamic NHSE. Our findings provide valuable insights into the underlying physics, facilitating a deeper comprehension of non-equilibrium phases.

Recently, experimental observations have revealed dynamic NHSE[48]. However, the realization of numerous unconventional non-Hermitian dynamics such as the chiral Zener tunneling, anharmonic Rabi oscillations, and wave self-acceleration remains elusive in experiments. Optical [49], acoustic [50,51], and mechanical systems [52] hold great promise as experimental platforms for observing these phenomena and validating our methodologies and conclusions. Our research provides theoretical support for the manipulation, trapping, and amplification of wave propagation, which can be readily implemented in classical experimental setups.

## ACKNOWLEDGMENTS

This work was supported by the National Key R&D Program of China (Grant No. 2022YFA1404400), the National Natural Science Foundation of China (Grant Nos. 12125504 and 12074281), and the "Hundred Talents Program" of the Chinese Academy of Sciences.## REFERENCES

[1] Bender, C. M. & Boettcher, S. Real spectra in non-Hermitian Hamiltonians having PT symmetry. Phys. Rev. Lett. **80**, 5243–5246 (1998).

[2] Bender, C. M. Making sense of non-Hermitian Hamiltonians. Rep. Prog. Phys. **70**, 947–1018 (2007).

[3] Yoshida, T., Peters, R. & Kawakami, N. Non-Hermitian perspective of the band structure in heavy-fermion systems. Phys. Rev. B **98**, 035141 (2018).